\documentclass[a4paper,oldversion,letter]{aa} 
\usepackage{graphicx}
\usepackage{txfonts}
\begin{document}
\def\cgs{erg cm$^{-2}$ s$^{-1}$}
\def\nh{cm$^{-2}$}
\def\lum{erg s$^{-1}$}
\def\gs{\hbox{\raise0.5ex\hbox{$>\lower1.06ex\hbox{$\kern-1.07em{\sim}$}$}}} 

   \title{The XMM Deep survey in the CDF-S I. First results on heavily obscured AGN\thanks{This work is based on observations obtained with XMM-{\it Newton}, an ESA science mission with instruments and contributions directly funded by ESA Member States and the USA (NASA).}}



   \author{A. Comastri         \inst{1}
          \and
          P. Ranalli \inst{2,1} 
           \and 
          K. Iwasawa \inst{3,1} 
            \and          
          C. Vignali \inst{2,1}
            \and
	  R. Gilli \inst{1}
	    \and
	  I. Georgantopoulos \inst{1}
	    \and
	   X. Barcons \inst{4}
	  \and 
          W.N. Brandt \inst{5,6} 
            \and
	  H. Brunner \inst{7} 
             \and
          M. Brusa \inst{7,8} 
           \and 
	  N. Cappelluti \inst{1,8} 
  	   \and 
	  F.J. Carrera \inst{4} 
           \and 
          F. Civano \inst{9}
           \and
          F. Fiore \inst{10}
	  \and
	  G. Hasinger \inst{11}
	  \and
	  V. Mainieri \inst{12}  
	  \and
	  A. Merloni \inst{7} 
	  \and 	
	  F. Nicastro \inst{10,9} 
	  \and 	
	  M. Paolillo \inst {13} 
	  \and 	
	  S. Puccetti \inst{14}
 	  \and 	
	  P. Rosati \inst{11}  
	  \and 
          J.D. Silverman \inst{15} 
	  \and
	  P. Tozzi \inst{16} 
	  \and 
	  G. Zamorani \inst{1} 
	  \and
	  I. Balestra \inst{7} 
	  \and
	  F.E. Bauer \inst{17} 
	  \and
	  B. Luo \inst{5}
	  \and
	  Y.Q. Xue \inst{5} 
          }

   \offprints{andrea.comastri@oabo.inaf.it}

   \institute{ INAF --  Osservatorio Astronomico di Bologna, via Ranzani 1
 I-40127 Bologna, Italy \email{andrea.comastri@oabo.inaf.it}
	\and
  Dipartimento di Astronomia, Universit\`a di Bologna, via Ranzani 1
 I-40127 Bologna, Italy 
	\and
 ICREA and Institut de Ci\`encies del Cosmos, Universitat de Barcelona (IEEC-UB),
  Mart\'i i Franqu\'es 1, E-08028 Barcelona, Spain
	\and
       Instituto de F\'\i{}sica de Cantabria (CSIC-UC), Avenida de los Castros, E-39005 Santander, Spain
    	\and 
  Department of Astronomy and Astrophysics, 525 Davey Lab, the Pennsylvania State University,
  University Park, PA 16802, USA
	\and
  Institute for Gravitation and the Cosmos, The Pennsylvania State University,
  University Park, PA 16802, USA 
	\and
 Max-Planck-Institut f\"ur extraterrestrische Physik,
  Giessenbachstrasse, 1, D-85748, Garching bei M\"unchen, Germany 
         \and
   University of Maryland, Baltimore County, 1000 Hilltop Circle, Baltimore, MD 21250, USA 
	 \and
       Harvard-Smithsonian Center for Astrophysics, 60 Garden
 Street, Cambridge, MA 02138, USA  
	\and
     INAF -- Osservatorio Astronomico di Roma, via Frascati 33
  I-00040 Monteporzio Catone, Italy
	\and
   Max-Planck-Institut f\"ur Plasmaphysik, Boltzmannstrasse 2, 85748 Garching bei M\"unchen, Germany
	\and
     European Southern Observatory, Karl-Schwarzschild-Strasse 2,
    D-85748, Garching bei M\"unchen, Germany 
	\and
   Dipartimento di Scienze Fisiche, Università di Napoli "Federico II," V. Cinthia, 9, 
  I-80126, Napoli, Italy 
	\and
     ASI Science Data Center, via Galileo Galilei,
       I-00044 Frascati, Italy
	\and
     IPMU, 	
     University of Tokyo, Kashiwanoha 5-1-5, Kashiwa, Chiba 277-8568, Japan 
	\and
	INAF --  Osservatorio Astronomico di Trieste, via Tiepolo 11, 
	I-34143 Trieste, Italy 
	\and
      Pontificia Universidad Cat\'olica de Chile, Departamento de Astronomía 
y Astrof\'isica, Casilla 306, Santiago 22, Chile 
 }
 
   \date{Received ; accepted }

\abstract{
We present the first results of the spectroscopy of distant, obscured AGN 
as obtained   with the ultra--deep ($\approx$ 3.3 Ms) XMM--{\it Newton} survey in the 
Chandra Deep Field South (CDF--S). One of the primary goals of the project 
is to characterize the X--ray spectral properties of obscured and heavily 
obscured Compton--thick AGN over the range of redhifts and luminosities 
that are relevant in terms of their contribution to the X--ray background.
The ultra--deep exposure, coupled with the XMM detector's spectral throughput, 
allowed us to accumulate good quality X--ray spectra 
for a large number of X--ray sources and, in particular, for heavily 
obscured AGN at cosmological redshifts. 
Specifically we present the X--ray spectral properties 
of two  high--redshift -- $z$= 1.53 and $z$=3.70 -- sources.
The XMM spectra of both are very hard, with a strong iron $K\alpha$ line 
at a rest--frame energy of $\sim$ 6.4 keV. 
A reflection--dominated continuum provides the best description of the 
X--ray spectrum of the $z=1.53$ source, while the intrinsic continuum 
of the $z=3.70$ AGN is obscured by a large column 
$N_H \approx$ 10$^{24}$ cm$^{-2}$ of cold gas. 
Compton--thick absorption, or close to it, is unambiguously detected in both 
sources. Interestingly, these sources would not be selected  
as candidate Compton thick AGN by some multiwavelength selection criteria based on the 
mid--infrared to optical and X--ray to optical flux ratios.}


    \authorrunning{Comastri et al.}
    \titlerunning{XMM-CDF--S}

     \maketitle                                                                
%

\section{Introduction}

The deepest X--ray surveys ever performed with the {\it Chandra} 
satellite in the Chandra Deep Field North (CDF--N, Alexander et al. 2003) 
and Chandra Deep Field South (CDF--S; Luo et al. 2008) have reached 
extremely faint X--ray fluxes ($\approx 2 \times 10^{-17}$ \cgs\ 
in the 0.5--2 keV band) and X--ray source surface densities 
of $\approx$ 10,000 per square degree. 
Both ultra--deep {\it Chandra} fields cover relatively small areas of the 
sky  (about 440 arcmin$^2$ each), and the majority of the sources  
are detected with a number of counts that prevents a detailed X--ray 
spectral analysis.

The ultra--deep XMM--{\it Newton} survey, with a nominal exposure of almost 3 Msec 
over an area encompassing the CDF--S and most of its four flanking fields 
(E--CDF--S), was conceived and planned to increase the counting statistics for 
a large number of X--ray sources, thanks to the large 
spectral throughput of XMM--{\it Newton}, 
and thus complements the {\it Chandra} and multiwavelength surveys information in 
that portion of the sky. 

The primary goal of the XMM--{\it Newton} ultra--deep survey in the CDF--S 
is the study of the X--ray spectral properties and cosmological evolution 
of heavily obscured and  Compton--thick ($N_H \ \gs \ 10^{24}$~cm$^{-2}$, 
hereafter CT; see Comastri 2004 for a review) AGN.
In recent years it has become clear that nuclear 
obscuration may be associated with  a specific, early phase in the 
cosmic history of every AGN (e.g. Page et al. 2004; Hopkins et al. 2006, Menci et al. 2008). 
At high redshift, the large gas reservoir may be able to both 
fuel and obscure the accreting supermassive black hole (SMBH) and sustain 
vigorous star formation in the host galaxy.
A sizable population of heavily obscured CT AGN at the redshift peak of 
AGN activity ($z\sim$ 1--2) is also invoked to reconcile the 
local SMBH mass function with  that obtained from the AGN luminosity 
function (Marconi et al. 2004).

The space density of obscured AGN beyond the local Universe is estimated 
by subtracting the cumulative flux and spectrum of 
resolved sources from the total X--ray background (XRB) 
intensity (Worsley et al. 2004; Comastri et al. 2005)
or in the framework of population synthesis models for the XRB
(e.g. Gilli et al. 2007; Treister \& Urry 2005). 
Following the approach described above, the most promising candidates to 
explain the so far largely unresolved spectrum of the XRB 
around its 20--30 keV peak are heavily 
obscured ($N_H >$10$^{23}$~cm$^{-2}$) AGN at $z$$\sim$0.5--1.5, many of them 
being CT.
The integral contribution of CT AGN to the hard ($>$ 10 keV) 
XRB and in turnto the cosmic SMBH accretion history depends
on the assumptions made in the synthesis models and ranges 
from about 10\% (Treister et al. 2009) up to $\sim$ 30\% (Gilli et al. 2007).
These estimates are made under simplified hypotheses, because the 
lack of observational evidence about the
cosmological evolution and the spectral shape of the most obscured AGN.
It is customary to constrain the CT AGN local space density 
by requiring that it matches the results obtained for optically 
selected 
(Risaliti et al. 1999; Guainazzi et al. 2005) or hard X--ray selected  
Swift (Tueller et al. 2010) and {\sc integral} (Beckmann et al. 2009) surveys
in the local Universe.
At present, only a handful of the ``bona fide" known CT AGN (Comastri 2004; 
Della Ceca et al. 2008) have been found at cosmological  distances (Norman et al. 2002; 
Iwasawa et al. 2005; Alexander et al. 2008)  
by means of X--ray observations. A systematic search for X--ray selected CT AGN 
has been carried out in the CDF--S
(Tozzi et al. 2006) and CDF--N (Georgantopoulos et al. 2009) yielding 
several candidates spanning a broad redshift range (up to $z$=3.7).
The surface density of the candidate CT AGN agrees fairly well 
with the Gilli et al. (2007) XRB synthesis model predictions.
Though effective in finding the signature of heavy obscuration, 
{\it Chandra} surveys are limited in photon--counting statistics.
The CT nature of most of the candidates is inferred on the basis 
of relatively poor quality (less than a few hundred  counts) X--ray spectra.
For this reason, alternative, multi--wavelength selection techniques 
based on high--ionization, narrow optical emission lines 
(Vignali et al 2010; Gilli et al. 2010) or on the ratio between mid--infrared 
and optical fluxes (Daddi et al. 2007; Alexander et al. 2008; Fiore et al. 2008 2009) 
were developed in the last few years.
They seem to be promising to find sizable samples of highly obscured and/or 
CT AGN at moderate to high redshifts ($z \sim$ 1--3). 
However, they rely on indirect methods, such as stacking, and  may still be 
prone to contamination by less obscured AGN or galaxies (e.g. Donley et al. 2008; 
Georgakakis et al 2010).

The ultra--deep XMM--{\it Newton} exposure over an area 
covering the CDF--S and its flanking fields coupled with the 
spectral throughput of the {\it pn} and MOS detectors is expected to deliver
X--ray spectra of unprecedented quality and thus to unambiguously identify 
CT absorption signatures in a significant number of objects.
In this paper (Paper I) we give an overview of the 
XMM--{\it Newton} survey in the CDF--S field and present the X--ray 
spectral analysis of two  obscured AGN.
We adopt a cosmology with $H_0$ = 70 km s$^{-1}$ Mpc$^{-1}$, $\Omega_M$=0.3, 
$\Omega_{\Lambda}$ = 0.7.

\vspace{-0.6cm}
\section{Observations and data analysis}

The CDF--S area was surveyed with XMM--{\it Newton} in several different epochs 
spread over almost nine years.
The results presented in this paper are obtained combining the observations 
awarded to our project in AO7 and AO8 (and performed in four different epochs between 
July 2008 and March 2010, with the archival data obtained in the period July 2001 -- 
January 2002).
The total exposure after cleaning from background flares is  $\approx$ 2.82 Ms for the 
two MOS and $\approx$ 2.45 for the {\it pn} detectors.
An extended and detailed  description of the full data set including 
data analysis and reduction and the X--ray catalog will be published
in Ranalli et al. (in preparation).

The data reduction was carried out with the standard XMM--{\it Newton} data analysis 
software SAS 9.0 and HEASARC's FTOOLS. 
The event files used for extracting spectra were filtered by applying light curves 
of the whole field in the high--energy band where few source photons are 
expected: 10--12 keV for the {\it pn} and 9.5--12 keV for the MOS, and the 
quiescent time intervals were selected. 
Background flares were filtered with a 3$\sigma$ clipping procedure.
Individual pointings were slightly off-set from each other to 
smear detector gaps and obtain a more uniform 
coverage of the field. 
{\it pn} observations are more affected by 
``high" particle background periods and detector gaps than MOS;
as a consequence, 
the net {\it pn} exposure time  is significantly shorter than 
for MOS. 
The individual pointings were brought to a common reference frame 
using the positions of the {\it Chandra} sources (Luo et al. 2008).
X--ray images in soft (0.4--1 keV), medium (1--2 keV) and hard (2--8 keV) 
bands were accumulated for each orbit and summed.
The color image is shown in Figure~1. 

\begin{figure} \includegraphics[width=0.49\textwidth]{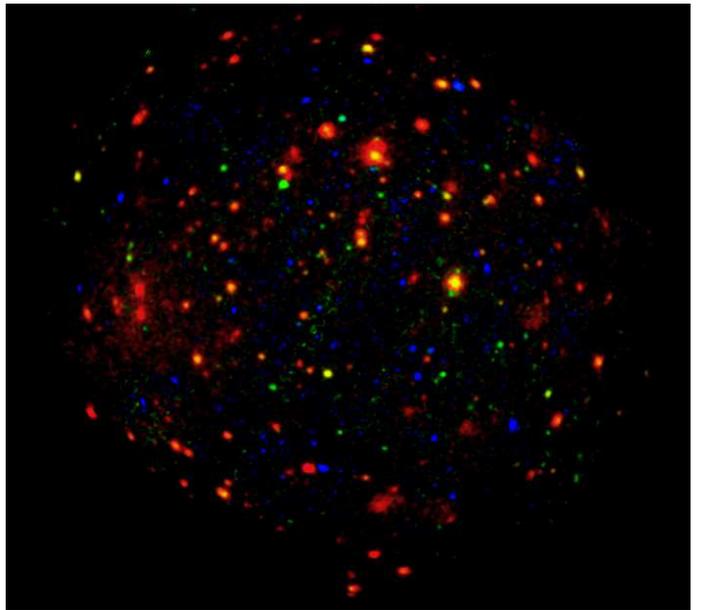} 
\caption{Combined {\it pn} and MOS color image. The standard color--coding is 
adopted : red for the soft (0.4--1 keV) band, green for the medium (1--2 keV) 
band and blue for the hard (2--8 keV) band. The energy range around the strong 
copper line at 7.8 keV in the {\it pn} data is removed. The image size is $\approx$ 
$32^{\prime}$ on a side and has been smoothed with a 4 arcsec Gaussian kernel.}
\label{Fig1} 
\end{figure}

%

\begin{figure*}[t]
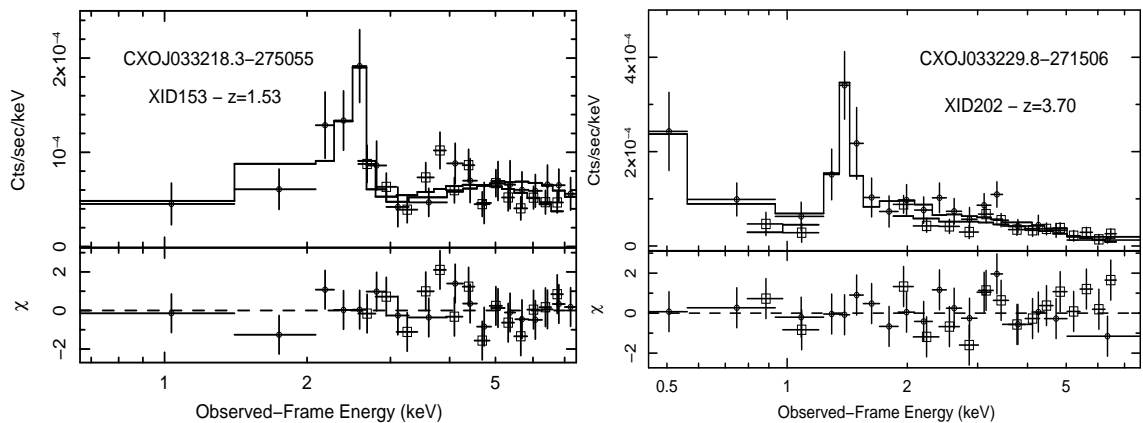

\begin{center}
\includegraphics[keepaspectratio=false, height=7.5cm, width=5.5cm,
angle=270]{16119fig2a.ps}
\includegraphics[keepaspectratio=false, height=7.5cm, width=5.5cm,
angle=270]{16119fig2b.ps}
\end{center}
\caption{Best--fit spectra and residuals of  XID 153 (left panel) and XID 202
(right panel). {\it pn} and MOS data are plotted as circles and squares, respectively. 
The spectra were mildly rebinned for plotting purposes.}
\end{figure*}

\section{X--ray spectral properties of two high redshift obscured AGN}

A sample of 14 candidate CT AGN from the CDF--S {\it Chandra} 1 Ms 
observations 
was selected by Tozzi et al. (2006) on the basis of a flat hard X--ray spectrum
typical of reflection--dominated sources. The presence of a strong 
(EW of the order of 1 keV) iron K$\alpha$ line 
is considered as additional evidence of heavy, Compton--thick obscuration. 

We present the XMM--{\it Newton} spectral analysis of two objects 
from the parent sample described above:
CXOCDFSJ033218.3-275055 at $z$=1.53 and CXOCDFSJ033229.8-275106 at $z=$3.7 
(XID 153 and XID 202, respectively, in the numbering scheme 
of Giacconi et al. 2002).
The former is the brightest X--ray source of the {\it Chandra}
sample of candidate CT AGN, while the latter is the highest redshift heavily obscured AGN 
known to date in the CDF--S (Norman et al. 2002).
\par
For each individual XMM orbit, source counts were collected 
from a circular region of 15 arcsec  radius, centered on the source position. 
Local background data were taken from a nearby region, where no source is found 
in the full exposure image or in the {\it Chandra} 2 Ms observations. 
The area of the background region is
larger than the source region by a factor of 1.7. The spectral data 
from individual exposures were summed up for the source and background, 
respectively, and the background subtraction was made assuming the common 
scaling factor for the source/background geometrical areas. 
{\it pn} spectra are extracted in the 0.5--7.8 keV energy range to avoid 
the strong $\sim$ 8 keV Cu line. Above $\sim$ 8 keV, the signal--to--noise ratio
decreases rapidly for relatively faint sources.
MOS spectra are extracted in the 0.5--9 keV range. X--ray photons in the 
1.2--1.8 keV range in the MOS 
are discarded to avoid instrumental background features and in particular the 
$\approx$ 1.5 keV Al line. 
MOS1 and MOS2 spectra are summed.
Response and effective area files were computed by averaging the individual files 
for each orbit. 
{\it pn} and MOS spectra were rebinned requiring at least 30 net counts per channel
and fitted simultaneously leaving the normalization 
free to vary. The {\it pn} count spectra are shown in Figure~2.
There is no clear evidence of significant flux or spectral  
variability between observations taken a few years apart. Variability studies on shorter 
timescales are hampered by the limited counting statistic.

Spectral fits are first performed with a single power law plus a Gaussian line. 
This modeling is purely phenomenological
and meant as a guide for more complex fits, which are summarized in Table 1.
We consider either a reflection--dominated continuum or a power law 
plus intrinsic absorption, partially or totally covering the central source.
In all fits a narrow (below the intrinsic instrumental energy resolution) 
iron line at $\sim$ 6.4 keV is included. Errors are quoted at the 1$\sigma$
confidence level for one parameter of interest.

\subsection{CXOCDFSJ033218.3-275055 (XID 153) $z$=1.53} 

The {\it pn} exposure time is about 1683 ks, after flare screening 
while the average MOS exposure time is about 2618 ks. The net counts in the 0.6--7 keV 
energy range are $\approx$ 1480. 
A fit to the joint {\it pn} and MOS spectra with a single power law returns 
an extremely hard ($\Gamma = -0.44\pm0.13$) continuum slope.  No useful spectral 
data can be obtained for the MOS below $\approx$ 2.5 keV. 
Line emission is clearly present 
(Fig.~2) and is best fitted by a neutral  (E=$6.35\pm0.04$ keV), 
strong (rest--frame equivalent width of 2330$\pm$520 eV) iron K$\alpha$ line. 
The fit quality is statistically acceptable ($\chi^2/dof$=36.7/39).
A reflection--dominated continuum ({\it pexrav} model in {\it xspec} assuming 
a face on geometry for the reflector with respect to the illuminating source, a cut--off 
energy of 200 keV for the primary continuum, and solar abundances) 
plus a Gaussian line well describes the observed spectrum.
The intrinsic power law slope ($\Gamma\sim$ 1.5) is within the range typical of Seyfert 
galaxies (i.e. Nandra et al. 1994). 
The iron line best--fit energy ($\simeq$ 6.4 keV) and equivalent width 
($\simeq$ 1.16 keV) are typical of reflection from cold matter (Table~1). 
\par
Even though a statistically acceptable fit ($\chi^2 = 32.7/38$) is obtained
with an absorbed power law ({\it plcabs} model in {\it xspec}), 
the continuum slope is extremely hard (see Table~1)  
and unphysical. The flat slope, coupled with the lack of a significant
photoelectric cut--off, mimics a reflection--dominated spectrum, which is then considered
the best--fit solution. There is no need for additional components to model the 
soft X--ray continuum (see Fig.~ 2 and $\S$ 3.2). 
The observed 2--10 keV X--ray flux is about 6.1 $\times 10^{-15}$ \cgs, and the corresponding 
luminosity in the 2--10 keV rest--frame is $\approx 1.8 \times 10^{43}$ erg s$^{-1}$.

\subsection{CXOCDFSJ033229.8-275106 (XID 202) $z$=3.70}

The {\it pn} exposure time is about 1892 ks, after flare screening 
while the average MOS exposure time is about 2720 ks. The source net 
counts in the 0.4--7.5 keV energy range are $\approx$ 1340.
A flat slope ($\Gamma=0.42\pm0.10$) is obtained with a 
single power--law fit to the joint {\it pn} and MOS spectra.
A strong (EW=1360$\pm$400 eV) iron line is clearly detected at a rest frame energy consistent 
with neutral or mildly ionized iron ($E=6.55\pm0.13$ keV).
The fit quality is marginally acceptable ($\chi^2/dof$=69.7/44) and is not improved 
by more complex models for the continuum such as a reflection--dominated 
spectrum or an absorbed power law (Table~1).
In both cases, positive residuals at low energies are present (Fig.~2). 
If a soft component, modeled with a steep power law, is added to 
a reflection dominated model for the hard continuum, the fit quality is only 
marginally improved (see Table~1). On the other hand, a remarkably good fit  
($\chi^2$/dof = 47.1/41) is obtained with a steep spectrum at low 
energies  plus an absorbed 
power law at high energies. The best--fit column density of the nuclear, 
hard X--ray continuum 
is on the order of  10$^{24}$ cm$^{-2}$. The soft component 
accounts for about 0.1\% of the unobscured flux at 1 keV.
The improvement with respect to an absorbed power law fit 
without a soft component is highly significant ($>$99.999\%) according to an F--test.
Even though the F--test cannot be used to assess the fit quality of models that
are not nested, the large difference ($\Delta\chi^2 \simeq$ 20--25 for 
1 to 3 degrees of freedom depending form the considered model in Table~1) 
in the $\chi^2$ statistic allows us to conclude 
that an absorbed power law plus a soft component is the best--fit solution.
To put our findings on solid grounds, given the relatively low 
signal--to--noise ratio of the observed spectrum, we performed 
some additional spectral fits. In particular, the parameters of the 
{\tt pexrav} model  are left free to vary in the reflection--dominated 
plus soft--component fits. The iron abundance and inclination angle
of the reflector turned out to be unconstrained by the present data.
The fit quality is improved ($\chi^2/d.o.f \approx$ 51/41); however, the 
best--fit parameters are very extreme: $\Gamma\approx 0.58$, $E_{cut}\approx$ 
10 keV. The flat power--law slope and the low cut--off energy 
mimic a peaked hard X--ray spectrum with a shape similar to an absorbed 
power law, thus reinforcing our choice of a heavily absorbed power law
as the best--fit model.  
The observed 2--10 keV flux is $\approx$ 3.3 $\times 10^{-15}$ \cgs.
Assuming the--best fit parameters (absorbed power law plus soft component), the 
intrinsic  nuclear luminosity in the 2--10 keV rest--frame 
is $\approx$ 6 $\times 10^{44}$ erg s$^{-1}$.

\begin{figure} 
\includegraphics[keepaspectratio=false,width=4cm,angle=0]{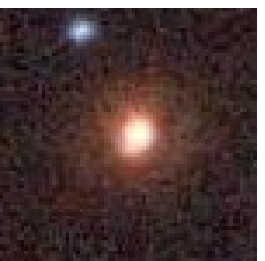}
\hfill
\includegraphics[keepaspectratio=false,width=4cm,angle=0]{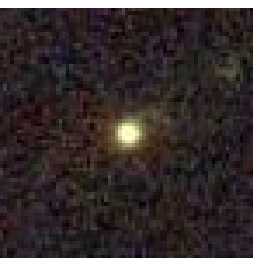}
\caption{Color--composite cutouts (2.2 arcsec across, 0.03 arcsec/pixel)
from the $b,v,i,z$ HST/ACS observations of GOODS--South 
(Giavalisco et al. 2004). {\it Left} XID 153. {\it Right} XID 202.} 
\label{Fig2} 
\end{figure}

\section{Discussion}

The counting statistics of the XMM spectra of the two candidates, high--redshift 
CT AGN analyzed in this work is of a quality 
to permit the determination of the presence of heavy absorption in both of them.
The relatively bright source at $z$=1.53 (XID 153) is best described by a 
reflection--dominated 
continuum and could be considered as a high--redshift analog of NGC 1068, 
the Seyfert 2 prototype in the local Universe. 
The intrinsic X--ray continuum of the $z$=3.70 Type 2 QSO (XID 202) is piercing through 
a high column density ($\sim 10^{24}$ cm$^{-2}$) absorber.
The very existence of heavily obscured and CT AGN at high redshift, predicted by 
AGN synthesis models for the XRB and theoretical models for AGN evolution, is confirmed
by these deep XMM--{\it Newton} observations of the CDF--S. 
Both sources are detected with a number of counts ($\approx$ 1300--1500) significantly 
higher than in the 2 Ms {\it Chandra} observations: about 200 net counts for XID 202 and
about 400 for XIS 153 (Bauer et al. in preparation). 
Also the combined {\it pn} and MOS 
signal--to--noise ratio of the XMM--{\it Newton} data ($\approx$ 23 and $\approx$19 
for XID 153 andXID 202 respectively)  is better than for {\it Chandra} data ($\approx$ 18 
and $\approx$14 for XID 153 and XID 202 respectively), despite the much lower level 
of the {\it Chandra} background. A comparable S/N will be achieved with the 4 Ms exposure.

The superior counting statistic of XMM data has allowed us to place significantly 
better constraints on the continuum spectral shape and absorption column densities 
of the two sources discussed in this paper. While their CT nature was only suspected 
on the basis of {\it Chandra} observations, thanks to the deep XMM--{\it Newton} 
observations it is now possible to unambiguously confirm the presence of CT 
circumnuclear matter in both of them 
and to establish whether the observed X--ray emission is caused by 
transmission or reflection.
\par
A self--consistent calculation of Compton--thick absorption and 
reprocessing features on the X--ray spectra has been recently published 
by Murphy \& Yaqoob (2009). The authors highlight the need for a proper treatment 
of absorption and scattering effects when the obscuring gas column density 
exceeds values on the order of a few 10$^{23}$ cm$^{-2}$. They also caution 
against an erroneous interpretation of the parameters obtained by approximate 
modeling of absorption by CT matter.  
We have performed spectral fits with the {\sc MYTorus} model\footnote{http:://www.mytorus.com}
routines  available on line (Murphy \& Yaqoob 2009). 
The best--fit column density of XID 202 
($\approx 1.0^{+0.2}_{-0.1} \times 10^{24}$ cm$^{-2}$) 
is perfectly consistent within the errors
with that obtained in Table~1. As far as the reflection--dominated fit of XID 153 
is concerned, we note that the signal--to--noise of the present 
data does not allow us to distinguish between the 
shape predicted by the Murphy \& Yaqoob (2009) model and the adopted {\sc pexrav} 
approximation. 
\par
Both objects are optically classified as type 2 AGN on the basis of 
high--ionization narrow emission lines (Szokoly et al. 2004). 
Both sources lie in the inner region of the CDF--S area, where deep, 
multi--wavelength data are available. 
While the rest--frame UV continuum of XID 202 is very weak and flat, 
the optical--UV continuum of XID 153 is extremely red (R--K $\simeq$ 5.7; Vega magnitudes).   
The HST cutouts, obtained combining ACS imaging in the $b$, $r$, $i$ and $z$ filters, 
are reported in Figure~3. The lower redshift object is clearly resolved and shows 
a spheroidal morphology, while the $z=3.7$ AGN is pointlike.

It is interesting to compare 
the broadband properties of these X--ray confirmed CT AGN 
with some multiwavelength selection criteria proposed to identify 
heavily obscured AGN at high redshift. 
The mid--infrared (24$\mu$m) to optical (R band) flux ratio (MIR/O) of XID 153 is 
$\sim$ 70, while that of XID 202 is $\sim$ 200. These values are well below 
the threshold (MIR/O$>$ 1000) adopted by Fiore et al. (2008) 
to select high--$z$ candidate obscured AGN. 
Galaxies showing a mid--IR emission in excess of that expected based on 
the star--formation rates measured from other multiwavelength data have a very 
hard {\it Chandra} stacked spectrum and are likely to host CT nuclei (Daddi et al. 2007).
Even if a possible overestimate of the value of the star--formation rate from mid--IR 
is taken into account (Murphy et al. 2009), both sources would 
have been selected as mid--IR excess.
\par
Another way to select obscured AGN at relatively high--$z$  is based on the 
X--ray--to--optical flux ratio (X/O)\footnote{$X/O=logf_X + R/2.5 + 5.5$ 
(Hornschemeier et al. 2001)}. 
A value on the order of 1 or higher is known to be 
a reliable proxy of optical obscuration (i.e. Fiore et al. 2003). In both sources the measured
X/O ratio is relatively high ($\sim$ 0.74 and 0.86 for XID 153 and XID 202 respectively),
but well within the distribution  observed for X--ray--selected AGN in various 
surveys. 

The two sources presented here would not have been selected 
as candidate CT AGN by two out of the three multiwavelength criteria 
discussed above. 
It should be noted that the efficiency in finding heavily obscured and CT 
AGN with multiwavelength  techniques is very low. 
Donley et al (2008) pointed out that a selection based on the 
mid--IR to optical spectral shape is likely to include a significant 
fraction of dusty, star--forming galaxies.  
Recently Yaqoob \& Murphy (2010) questioned the use of the 
mid--IR vs X--ray luminosity ratio as a reliable proxy for the column density of 
obscuring gas. Finally, Silverman et al. (2010) 
suggest that most of the high X/O ratio sources in the CDF--S are optically obscured, but 
at moderate 0.5 $<z<$ 1.5 redshifts. 

The full exploitation of the ultra--deep XMM--{\it Newton} survey 
in the CDF--S will allow us to uncover the unambiguous signatures of 
heavy, Compton--thick absorption for a sizable sample of 
relatively X--ray bright sources and, thanks to the extensive multi--wavelength 
coverage of the field, constrain their broadband spectral energy 
distributions and validate or devise additional methods
for the selection of much fainter sources (Balestra et al., in preparation).

An important question that can be addressed 
thanks to deep X--ray observations is the distribution of obscuring gas 
column densities in the Compton--thick regime and, in particular, the relative 
fraction of objects for which the intrinsic nuclear continuum 
is able to pierce through the obscuring gas, which makes it 
possible to estimate their nuclear luminosity.
These sources are dubbed transmission--dominated or ``mildly" Compton--thick
and are expected to contribute to the hard X--ray background emission.
The X--ray spectrum of reflection--dominated or ``heavily" CT AGN 
($N_H \gg 10^{24}$ cm$^{-2}$, $\tau \gg$ 1) is depressed by 
Compton down--scattering over the entire X--ray band.
As a consequence, their contribution to the XRB flux density is likely to be negligible 
with respect to that of mildly CT AGN. However, they might still be sufficiently numerous 
 to provide a non--negligible contribution to the SMBH mass density.

It is remarkable that previous searches in the CDF--S (Tozzi et al. 2006) 
and CDF--N (Georgantopoulos et al. 2009) seem to suggest that 
reflection--dominated sources are much more abundant than ``mildly" CT AGN 
(i.e. 9 out of the 10 candidates in the CDF--N and all 14 candidates 
in the CDF--S were considered to be reflection--dominated, though 
the present results indicate that XID 202 at least is not). However, 
given that CT candidates in the {\it Chandra} deep fields are identified 
on the basis of a flat hard X--ray spectrum, a strong bias toward reflection--dominated 
AGN is likely to be present in these {\it Chandra} samples.
We expect to detect and recognize about 30--40 CT AGN in the CDF--S. About 
30--40\% of them (Gilli et al. 2007) will be detected with a number of counts comparable 
to the two presented here.
The obscuring column density will be measured with an accuracy that is good enough 
to distinguish between heavy absorption and reflection.

\section{Conclusions} 

The spectral capabilities of the {\it pn} and MOS camera onboard XMM--{\it Newton} 
coupled with an ultra--deep exposure fully meet the expectation and are providing 
high quality X--ray spectra, which will allow us to address several 
important aspects concerning the physics and the evolution of the 
most obscured AGN. As an example of the quality of the future results that can be obtained 
with these data, we showed that

\begin{itemize}

\item[$\bullet$] The XMM--{\it Newton} deep exposure in the CDF--S allowed us 
to obtain good quality X--ray spectra of two high--redshift AGN, which are suspected to host 
a CT nucleus on the basis of {\it Chandra} observations.

\item[$\bullet$] The high--energy emission of the $z=3.7$ AGN is obscured by cold matter 
with a column density close to but slightly lower ($\tau\approx 0.7$) than
unit optical depth for Compton scattering.
Reflection from a Compton--thick reprocessor is the most likely explanation 
for the very hard X--ray spectrum of the $z=1.53$ source.

\item[$\bullet$] The comparison with some of the multiwavelength criteria, 
devised for the selection of Compton--thick AGN at high redshift, highlights
the importance of deep X--ray observations for a robust determination of 
the obscuring gas properties.

\end{itemize}

\begin{acknowledgements}
The authors would like to 
warmly thank Maria Diaz Trigo and Ignacio della Calle
for their valuable help and advice in the scheduling of the XMM--CDF--S observations.
Support from the Italian Space Agency (ASI) under the 
contracts ASI-INAF I/088/06/0 and I/009/10/0 is acknowledged. 
AC acknowledges the NASA grant NNX09AQ05G to visit CfA, 
where part of this work was written. 
IG and AC acknowledge the 
Marie Curie fellowship FP7-PEOPLE-IEF-2008 Prop. 235285.
XB and FJC acknowledge financial support from the Spanish
Ministerio de Ciencia e Innovaci\'on under project AYA2009-08059. 
MB acknowledges support by the Deutsches Zentrum f\"ur Luft- und Raumfahrt, DLR 
project numbers 50 OR 0207 and 50 OR 0405. 
FC, FN and MB acknowledge NASA grants NNX08AX51G and NNX09AQ05G.
FC acknowledges support from the Blanceflor Boncompagni Ludovisi Foundation.
JDS is supported by World Premier International 
(WPI) Research Center Initiative , MEXT, Japan. WNB acknowledges NASA grant 
SV0-80004 and NASA ADP grant NNX10AC99G. FEB cknowledges the support of CONICYT,
Chile, under grants FONDECYT 1101024 and FONDAP (CATA) 15010003.
\end{acknowledgements}

\footnotesize{
\begin{table*}
\caption{Best--fit parameters for reflection--dominated and transmission--dominated models}
\begin{tabular}{cccccccc} 
\hline\hline             
Source ID & Model$^a$ &$\Gamma$ & $\Gamma^b_S$ &  $N_H$ & $E_{K\alpha}$ & $EW^c$ & $\chi^2$/d.o.f  \\
          & &    &  & 10$^{22}$ cm$^{-2}$   & keV & eV &  \\
\hline
153      & {\bf REF} & 1.50$\pm$0.10 & ...  &  ...  & 6.36$^{+0.07}_{-0.05}$ & 1160$^{+440}_{-350}$ & 28.9/39  \\
153      & ABS & $-$0.11$\pm$0.22 & ...  &  5.0$^{+3.5}_{-2.1}$  & 6.35$\pm$0.04 & 1870$^{+410}_{450}$ & 32.7/38  \\
\hline\hline             
202      & REF & 1.99$\pm$0.10 & ... & ...  & 6.57$\pm$0.13 & 760$^{+340}_{-250}$ & 72.4/44  \\
202      & PL+REF & 1.95$\pm$0.10 & 7.3$^{+0.4}_{-0.6}$ & ...  & 6.57$\pm$0.13 & 930$^{+300}_{-330}$ & 60.9/42  \\
202      & ABS & 1.48$^{+0.33}_{-0.40}$  & ... & 58$^{+32}_{-22}$  &  6.54$\pm$0.15 & 840$^{+290}_{-470}$ & 69.4/43  \\
202      & {\bf PL+ABS} & 1.82$^{+0.34}_{-0.59}$ & 4.1$^{+0.5}_{-0.9}$ & 88$^{+68}_{-40}$ &  6.53$\pm$0.14 & 840$\pm$320 & 47.1/41  \\
\hline
\end{tabular}
\vspace{0.5 cm}
\par
Notes: \\
Errors are at 1$\sigma$. The best fit model is in {\bf boldface} in column 2. \\
{\bf a} Fitted models: reflection--dominated (REF), absorbed power--law (ABS), power law plus absorbed power--law (PL+ABS).\\
{\bf b} $\Gamma_S$ is the power--law slope of the soft component. \\
{\bf c} The iron line EW with respect to the intrinsic (unabsorbed) continuum is $\approx$ 1660 eV for XID 153 
and $\approx$ 160 eV for XID202. \\
\end{table*}}

\end{document}